\pdfoutput=1

\documentclass[aps,prb,reprint,showpacs,superscriptaddress]{revtex4-1}
\usepackage{graphicx}
\usepackage{graphics}
\usepackage{subfigure}
\usepackage{amsmath}
\usepackage{amssymb}
\usepackage{amsfonts}
\usepackage{dcolumn}
\usepackage{dsfont}
\usepackage{epstopdf}
\usepackage{latexsym}
\usepackage{rotating}
\usepackage{color}
\usepackage{latexsym}
\usepackage{bbm}
\usepackage{float}
\usepackage{epsfig}
\usepackage{epsf}
\usepackage{psfrag}
\usepackage{bm}
\usepackage{amsthm}
\usepackage{eucal}
\usepackage{mathrsfs}
\usepackage{url}
\usepackage{braket}
\usepackage{soul}
\usepackage[none]{hyphenat}
\usepackage{color} 

\usepackage{hyperref}
\hypersetup{
colorlinks=true,final=true,
        linkcolor=blue,
        citecolor=blue,
        filecolor=blue,
        urlcolor=blue,
}
\begin{document}
\title{Emerging topological states in EuMn$_2$Bi$_2$: A first principles prediction}

\author{Amarjyoti Choudhury}
\author{T. Maitra}
\email{tulika.maitra@ph.iitr.ac.in}
\affiliation{Department of Physics, Indian Institute of Technology Roorkee, Roorkee - 247667, Uttarakhand, India}
\date{\today}

\begin{abstract}
New materials with magnetic order driven topological phases are hugely sought after for their immense application potential. In this work, we propose a new compound EuMn$_2$Bi$_2$ from our first principles density functional theory calculations to host novel topological phases such as Dirac/Weyl semimetal and topological insulator in its different magnetic states which are energetically close to one another. We started with an isostructural compound EuMn$_2$As$_2$ where the magnetic structure has been studied experimentally. From our calculations we could explain the nature of two magnetic transitions observed experimentally in this system and could also establish the correct magnetic ground state. Our electronic structure calculations reveal the insulating nature of the ground state consistent with the experiments. By replacing all As by Bi in EuMn$_2$As$_2$ and by optimizing the new structure, we obtained the new compound EuMn$_2$Bi$_2$. We observe this compound to be dynamically stable from our phonon calculations supporting its experimental preparation in future. By comparing the total energies of various possible magnetic structures we identified the ground state. Though the magnetic ground state is found to be insulating in nature with tiny band gap which is an order of magnitude less than the same in EuMn$_2$As$_2$, there were other magnetic states energetically very close to the ground state which display remarkable non-trivial band topology such as Dirac/Weyl points close to the Fermi level and topological insulator state. The energetic proximity of these magnetic order driven topological phases makes them tunable via external handle which indicates that the proposed new material EuMn$_2$Bi$_2$ would be a very versatile magnetic topological material.       
  
\end{abstract}

\maketitle

\section{Introduction} 
The interplay between magnetism and band topology in quantum materials has created a huge surge in research activities of late\cite{stp2,stp3,stp4,WSMS,TI,DSMS,DSMS1,NSMS,NSMS1}. Several Eu based 122-Pnictide compounds have come to fore in this field of research for their emergent topologically protected states driven by magnetic order\cite{AA,soh,jyoti,Marshall,AAA,tm}. Particularly, EuCd$_2$As$_2$ has drawn a lot of attention for being identified as the first magnetic order driven ideal Weyl semimetal both theoretically as well as experimentally\cite{AA,soh,jyoti}. The magnetically driven topological states has been reported so far in the compounds where Eu is the only magnetic ion\cite{AA,tm}. However, in the compounds such as EuMn$_2$X$_2$ (X= As, Sb, P) where Mn is also magnetic in addition to Eu, no such topological states have been reported even though these compounds display very intriguing magnetic properties such as more than one magnetic phase transitions, spin reorientations etc.\cite{anand} 

\begin{figure}
	\begin{center}
		\includegraphics[width=5.5cm]{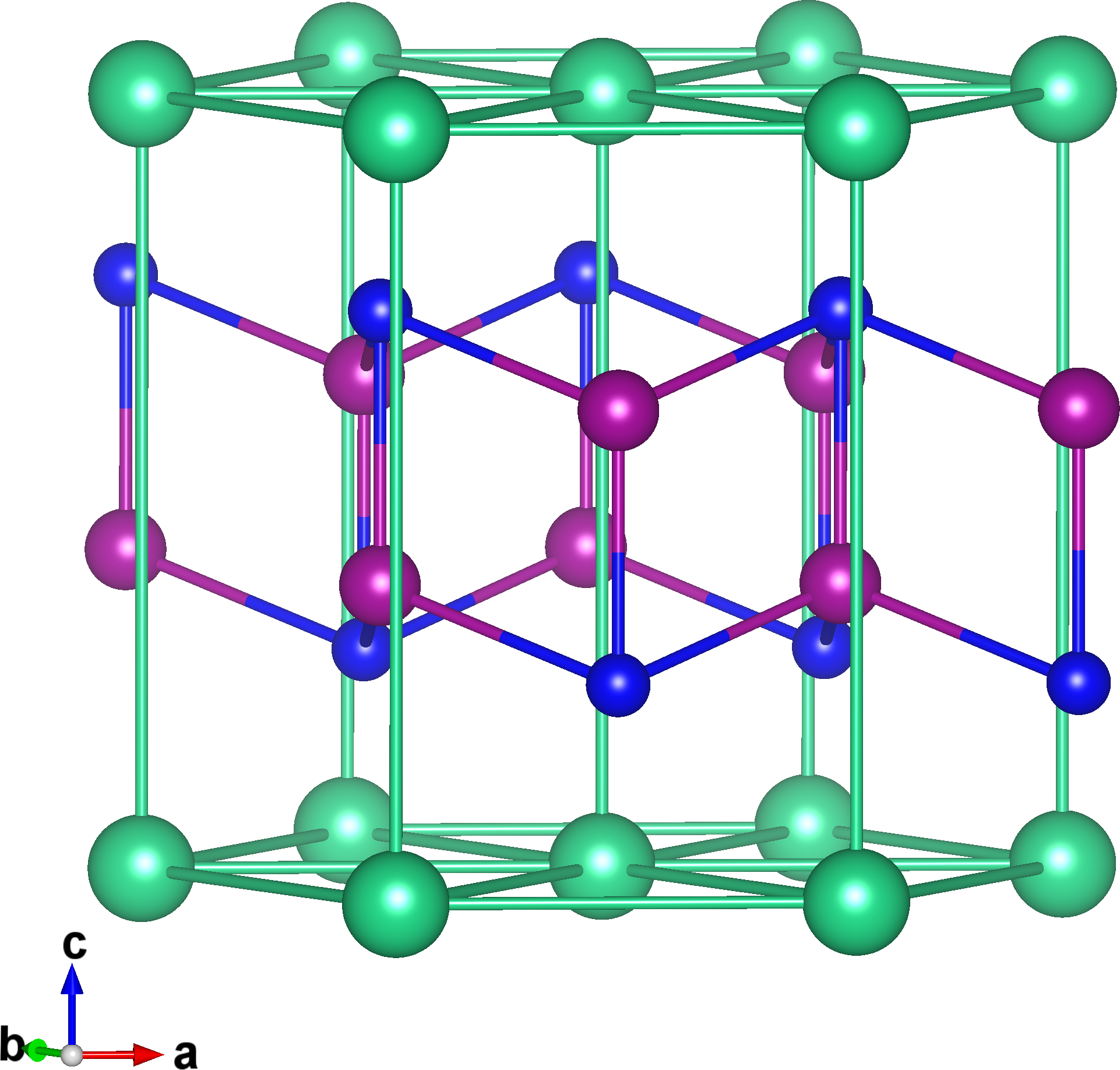}
		\caption{\label{Fig1} \noindent The crystal structure of EuMn$_2$X$_2$ (X=As/Bi) with space group P$\bar{m}$31 where Eu, Mn and As/Bi ions are shown in green, dark pink and blue colours respectively.}
	\end{center}
\end{figure}

122 Pnictides, typically, adopt a ThCr$_2$Si$_2$-type tetragonal structure\cite{1}. However, these Eu based 122 Pnictides stand out as exception to this common crystal symmetry. Instead, they crystalize in a trigonal structure known as the CaAl$_2$Si$_2$-type structure \cite{H.Zhang,C.Zheng}, with the P${\bar 3}$m1 space group. The crystal structure of EuMn$_2$As$_2$ is shown in  Fig~\ref{Fig1} where Eu$^{2+}$ ions with spin moment 7/2 form a triangular lattice layers separated along $c$ direction by Mn$_2$As$_2$ layers. Mn$^{2+}$ ions with spin moment 5/2, on the other hand, form corrugated honeycomb lattice with Eu ions sitting at center of the honeycomb when seen from the top. 
\begin{figure}
	\begin{center}
		\includegraphics[width=8.5cm]{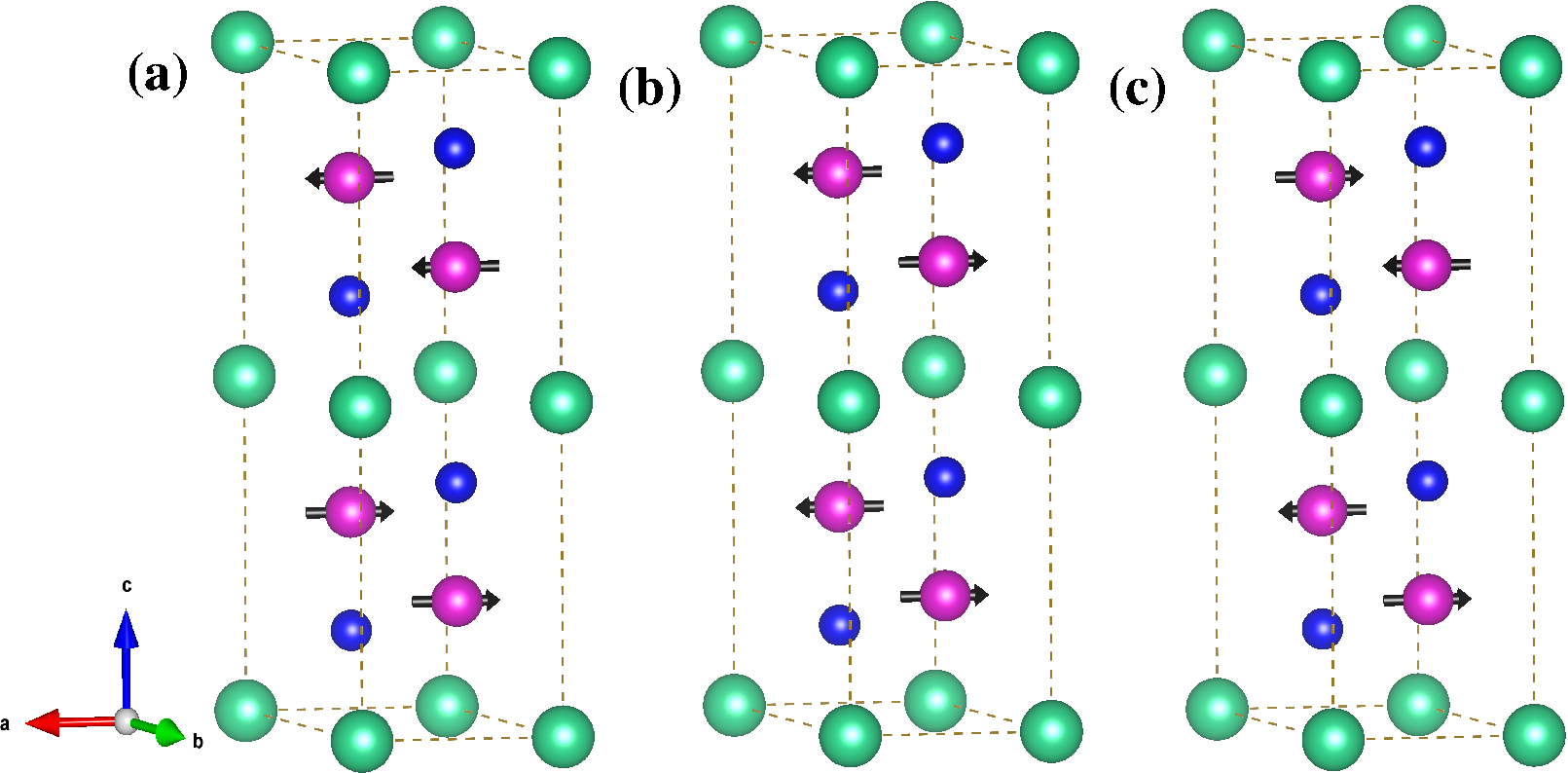}
	\end{center}
	\caption { \noindent Different magnetic configurations of Mn sublattice considered in our calculation (see text): (a) A0-AFM, (b) C0-AFM (c) G0-AFM.}
	\label{Fig2}
	
\end{figure}
V. K. Anand et.al \cite{anand} studied the magnetic and transport properties of EuMn$_2$As$_2$ and K doped EuMn$_2$As$_2$ in detail few years back. The authors reported the presence of two magnetic transitions, one at higher temperatures of around 142K (inferred from specific heat measurements) while the other at lower temperatures around 14K (from susceptibility measurements). The 142K transition was proposed to be associated with ordering of Mn spins while the 14K one was linked to the ordering of Eu spin moments\cite{anand}. In addition, they also indicated about a possible spin-reorientation transition at 5K. Though the presence of two magnetic transitions were reported by V. K. Anand et al., exact nature of magnetic ordering of Mn and Eu spins was not very clear. To probe this Dahal et al. studied the system using elastic neutron scatterring measurements\cite{dahal}. The authors presented compelling evidence of two distinct magnetic transitions; one at 135K (T$_{N1}$) and the other at 14.4K (T$_{N2}$). Subsequent numerical modeling of their neutron data yielded valuable insights into the nature of spin correlations within both the magnetic phases (T $<$ 14.4K and 14.4K $<$ T $<$ 135K). From their analysis, they concluded that the Mn spins order below 135K with C-type antiferromagnetic (C-AFM) order whereas Eu spins order below 14.4K with A-type antiferromagnetic (A-AFM) order. Both Mn and Eu spins display non-collinear magnetic ordering with spin moments pointing towards diagonal direction\cite{dahal}. Note that due to Eu's significant neutron absorption capacity, corrections were applied to the elastic neutron data while modelling. Regarding the transport properties of EuMn$_2$As$_2$, Anand et al.\cite{anand} observed it to be insulating in nature with activation energy of about 50 meV. They also observed that a very small amount of K doping at Eu sites leads to a metallic state\cite{anand}. An understanding of the underlying mechanism behind the multiple magnetic transitions, possibility of spin-reorientation and exact nature of magnetic orders in EuMn$_2$As$_2$ demands a thorough theoretical (particularly abinitio) investigation. More importantly, it would be worth exploring if magnetically driven topological states as seen in EuCd$_2$As$_2$\cite{AA} or EuMg$_2$Bi$_2$\cite{Marshall,tm,pakhira1} can be achieved in EuMn$_2$X$_2$ systems.    
\begin{table}
	\begin{center}
		\caption{The variation of total energy difference between   (G0-AFM / A0-AFM/ FM) and C0-AFM ($\Delta$E) in  meV with U$_{eff}$ in eV.} 
		\label{Tab1}
		\begin{tabular}	{m{6em} m{2.0cm} m{2.0cm} m{2.0cm}  } 
			\hline\hline \\
			& G$0$$-$C$0$ & A$0$$-$C$0$ & FM$-$C$0$ \\\\
			\hline \\
			{\bf GGA} & 29.72 & 533.71 &580.31 \\\\
			\hline \\
			{\bf  GGA+U} \\\\ U$_{eff}$(eV)  & &  &    \\ \\
			3   & 18.33 & 285.34 & 323.75 \\\\
			5   & 13.15 & 196.36 & 221.98 \\\\
			7   & 9.61 & 141.29 & 157.67 \\\\
			11  & 5.28 & 80.07 & 86.77 \\\\
			\hline
			
			\hline\hline
		\end{tabular}	
	\end{center}
\end{table}

In this work we have tried address the two important issues mentioned above. Firstly, we performed a thorough investigation of the nature of non-collinear magnetic ground state and if there is any spin re-orientation in EuMn$_2$As$_2$ using first principles density functional theory calculations. Secondly, we explored the properties of a new compound (not reported experimentally yet) EuMn$_2$Bi$_2$ by substituting As of EuMn$_2$As$_2$ by Bi. Remarkably, we find EuMn$_2$Bi$_2$ in its magnetic ground state to have a very tiny band gap, an order of magnitude less than the same in EuMn$_2$As$_2$. More importantly, certain low energy excited states with different magnetic orders or spin orientation are found to be topologically non-trivial. Slight change in the magnetic order of Eu or Mn sublattice or the spin orientation seems to be driving the system to a topological insulator $/$ Dirac (Weyl) semimetallic state with the presence of band inversion and interesting surface states near the Fermi level. 
\begin{table}
	\begin{center}
		\caption{The coordinates (k$_x$, k$_y$, k$_z$) of two pairs of WPs in the BZ for EuMn$_2$Bi$_2$ in C0-AFM1 and G0-AFM5 phases are given in the units of 1/\AA. Their relative energies with respect to Fermi level and chirality (C) are also mentioned.} 

		\label{Tab2}
		\begin{tabular}	{m{1em} m{0.8cm} m{0.8cm} m{1.0cm} m{1.5cm} m{1.0cm} m{1.0cm}   }
			\hline\hline \\
			& & &  C0$-$AFM1 \\
			\hline\\
			& &k$_x$& k$_y$&  k$_z$ & E-E$_f$ (eV)  & C  \\\\
			\hline
			&W$_{c1+}$ & 0. & 0  & $+$ 0.0624 &    0.042 & $-1$   \\\\
			&W$_{c1-}$ & 0 & 0  & $-$ 0.0624  & 0.042 &  $+1$ \\\\
			&W$_{c2+}$ &0  & 0  & $+$ 0.0623 &    0.042 & $+1$   \\\\
			&W$_{c2-}$ & 0 & 0  & $-$ 0.0623 &   0.042 & $-1$ \\\\
				\hline \\\\
			& & &  G0$-$AFM5 \\
						\hline\hline \\
			& &k$_x$& k$_y$&  k$_z$ & E-E$_f$ (eV)  & C  \\\\
			\hline \\
			&W$_{g1+}$ &0  & 0  & $+$ 0.097 &   0.319  & $-1$   \\\\
			&W$_{g1-}$ & 0 & 0  & $-$ 0.097 &   0.319 & $+1$ \\\\
			&W$_{g2+}$ &0  & 0  & $+$ 0.148  &    0.352 & $+1$   \\\\
			&W$_{g2-}$ & 0 & 0  & $-$ 0.148  &   0.352 & $-1$ \\\\
			
			\hline
			\hline \\\\
			
		\end{tabular}	
	\end{center}
\end{table}
\begin{figure*}[!htp]
	\begin{center}
		\includegraphics[width=17.0cm]{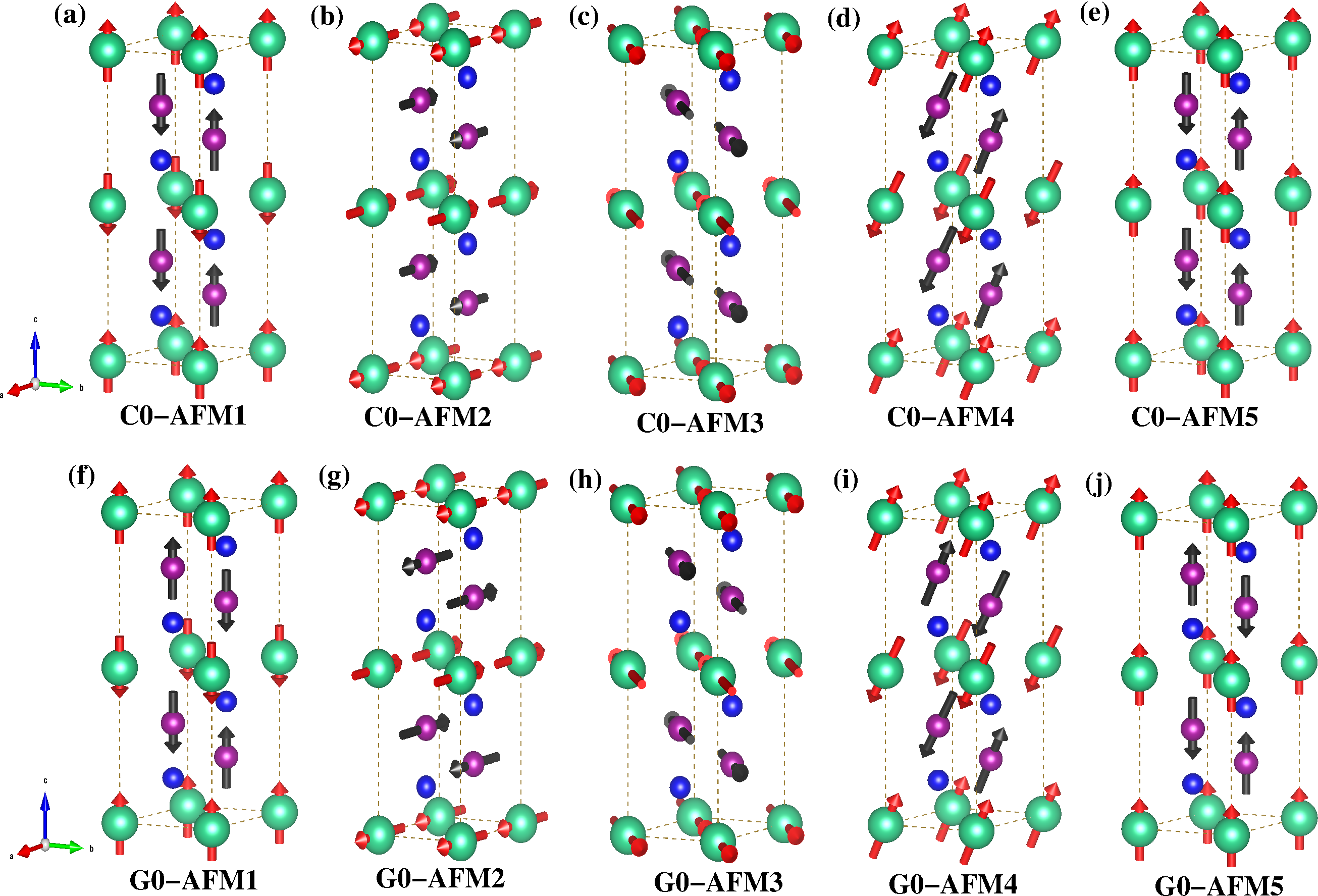}
	\end{center}
	\caption { \noindent Some of the magnetic configurations (where both Eu and Mn moments order) considered in our calculations are depicted. }
	\label{Fig3}
	
\end{figure*}
\section{Methods}
\label{meth}
We performed total energy and electronic band structure calculations for EuMn$_2$Bi$_2$ using DFT calculations with the Perdew-Burke-Ernzerhof generalized gradient approximation (PBE-GGA)\cite{PBE} exchange-correlation functional. The computations were carried out utilizing the projector-augmented wave (PAW)\cite{PAW} approach and a plane-wave basis set within the Vienna Ab-{\sl{initio}} Simulation Package (VASP).\cite{kresse, kresse2}In our calculations, we accounted for both Coulomb correlation and spin-orbit (SO) interaction due to the strong localization of 4f and 3d electrons in the Eu and Mn ions. In our self-consistent calculations, we utilized a (13x13x7) $\Gamma$-centered Monkhorst-Pack \cite{Mon} {\bf k} point mesh within the Brillouin zone (BZ) and set the kinetic energy cutoff for the plane wave basis set to 460 eV. In our GGA+U calculations, we followed the Dudarev formalism, \cite{Duda}  employing an effective Hubbard interaction denoted as $U_{\text{eff}} = U - J$, where $U$ represents Coulomb correlation and $J$ stands for Hund's exchange. We focused on the case with $U_{\text{eff}} = 7 \, \text{eV}$ and $U_{\text{eff}} = 5 \, \text{eV}$, particularly applied to the Eu 4f states and Mn 3d states. However, we also explored a range of values for $U_{\text{eff}}$, spanning from $3 \, \text{eV}$ to $11 \, \text{eV}$, to ensure the robustness and reliability of our outcomes. To analyze the topological characteristics of EuMn$_2$Bi$_2$, we utilized the Wannier90\cite{Mostofi} and WannierTools\cite{Wu} software packages. Wannier90 employs Maximally Localized Wannier Functions (MLWF)\cite{Vander} to construct a tight-binding model by fitting DFT bands. Subsequently, this tight binding model is used to compute various topological properties via WannierTools. 

\begin{figure}
	\begin{center}
		\includegraphics[width=8.5cm]{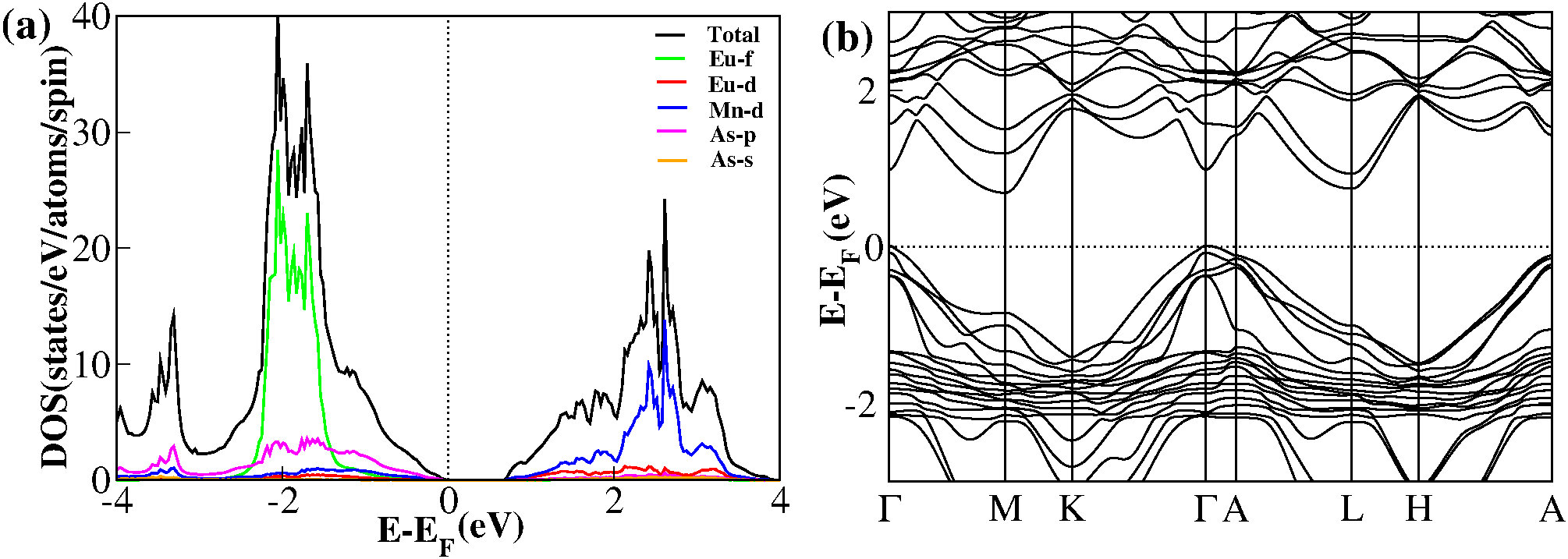}
	\end{center}
	\caption { \noindent (a) DOS (b) band structure of EuMn$_2$As$_2$ in it's magnetic ground state calculated within GGA+U+SO approximation. Insulating nature is clearly seen.}
	\label{Fig4}	
\end{figure}
\section{Results and Discussion}
\label{res}
\subsection{EuMn$_2$As$_2$}
{\it Ground state magnetic order and spin reorientation:} 
As discussed above EuMn$_2$As$_2$ undergoes two magnetic transition and it is believed that the higher temperature one at 135K is caused by the Mn spin ordering whereas the lower one around 14K is caused by the Eu spin ordering\cite{dahal}. In our endeavor to verify these claims and establish the cause of magnetic transitions from theoretical perspective, we undertook the total energy calculations for various magnetic configurations under different situations as discussed below. Firstly, to study the magnetic behaviour of the Mn sublattice we have performed total energy calculations considering Eu 4f electrons as core electrons so that there is no moment at Eu sites. We then calculated the total energies of various magnetic configurations which included ferromagnetic (FM) and distinct antiferromagnetic (AFM) configurations labeled as A0-AFM, C0-AFM, and G0-AFM, as depicted in Fig~\ref{Fig2}. These computations were performed under different approximations, such as GGA, GGA+U, and GGA+U+SO considering the optimized structural parameters. Within  GGA we observe the C0-AFM configuration to be the ground state (see Table~\ref{Tab1}) with an energy difference of approximately 0.5eV with FM and A0-AFM while G0-AFM is found to be only slightly higher (~0.02eV) in energy. We found C0-AFM to remain the ground state even after incorporating the Coulomb correlation (U) applied to the Mn d-states within GGA+U, though the energy difference between the G0-AFM and C0-AFM states is seen to diminish further (goes like 18.33 meV, 13.15 meV, 9.61 meV, and 5.28 meV for U$_{eff}$ values of 3 eV, 5 eV, 7 eV, and 11 eV, respectively). Finally, to find the preferred direction of the spin moments within the C0-AFM state we performed calculations for various spin orientations such as along the crystallographic $a$, $b$, $c$ and along [110], [111] etc. within  GGA+U+SO approximation. We observe that the Mn moments prefer to align between the crystallographic a and b axes (i.e. along [110]) as shown in Fig~\ref{Fig2}.

To investigate the magnetic ordering at lower temperatures (below 14K) where it is believed that in addition to Mn, Eu spins also order, we carried out calculations considering Eu 4f electrons as valence. Keeping the Mn spin moments in C0-AFM and G0-AFM (energetically the lowest two) arrangements, we calculated total energies for Eu spin moments in FM and A-AFM (as G-AFM and C-AFM are frustrated in triagular geometry) within GGA+U approximation. We observe that in the lowest energy state the Eu spins order in A-AFM which is consistent with the experimental observation\cite{anand, dahal}. Further to establish the preferred orientations of Eu and Mn spin moments, we performed total energy calculations within GGA+U+SO approximation with Eu and Mn both parallel to $c$, both parallel to $a$, Eu along $c$ and Mn along [110], Eu and Mn both along [111] direction totalling 32 different``non-collinear," ``non-coplanar," and ``non-collinear coplanar" magnetic configurations (see Table~\ref{Tab3} in Appendix-A and Fig~\ref{Fig3}). We observed that the total energy of the system is lowest when Eu and Mn both point along [110] direction as shown in Fig~\ref{Fig3}(c). However, the state with Eu/Mn moments pointing along [111] is only 0.04 meV higher in energy (see Fig~\ref{Fig3}(d)). Therefore, we do not observe any spin-reorientationin the ground state. However, our results suggest that with slight perturbation the spin moments can reorient to [111] direction. Further, our magnetic exchange interaction calculations suggest that the Mn sublattice will order at much higher temperatures ($\lvert$J$_{ab}$$\rvert$$\sim$ 0.78 meV and $\lvert$J$_{c}$$\rvert$$\sim$ 5.29 meV for Mn-Mn exchange) whereas Eu sublattice will order at a lower temperature ($\lvert$ J$_{ab}$$\rvert $ $\sim$ 0.02 meV and $\lvert$J$_{c}$$\rvert$$\sim$ 0.004 meV for Eu-Eu exchange) consistent with experimental observations.

\begin{figure}
	\begin{center}
		\includegraphics[width=8.5cm]{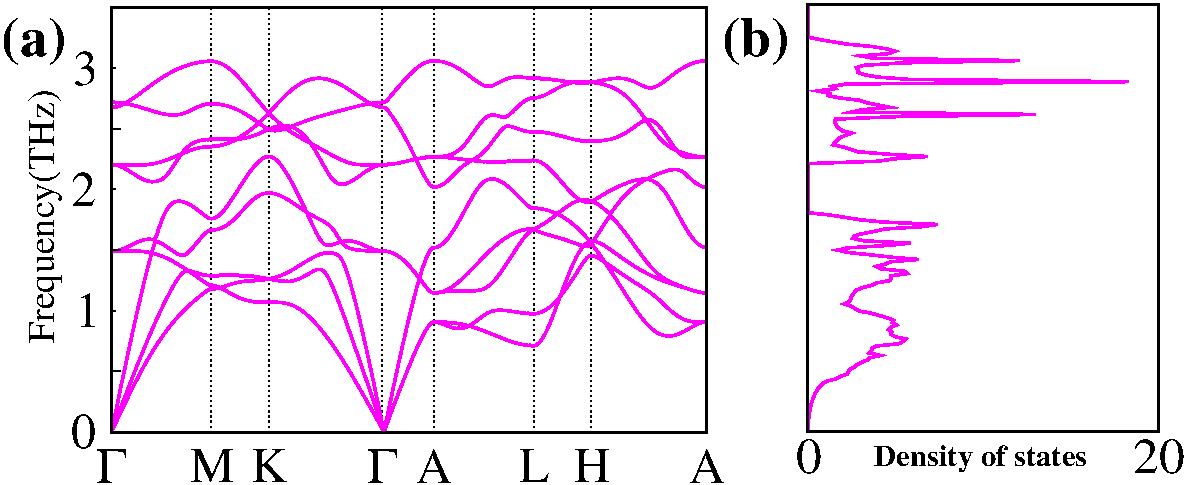}
		\caption{\label{Fig5} \noindent The phonon (a) band structure and (b) DOS of EuMn$_2$Bi$_2$ illustrating its dynamical stability. No soft phonon mode is observed.}
	\end{center}
\end{figure}
\begin{figure}
	\begin{center}
		\includegraphics[width=8.5cm]{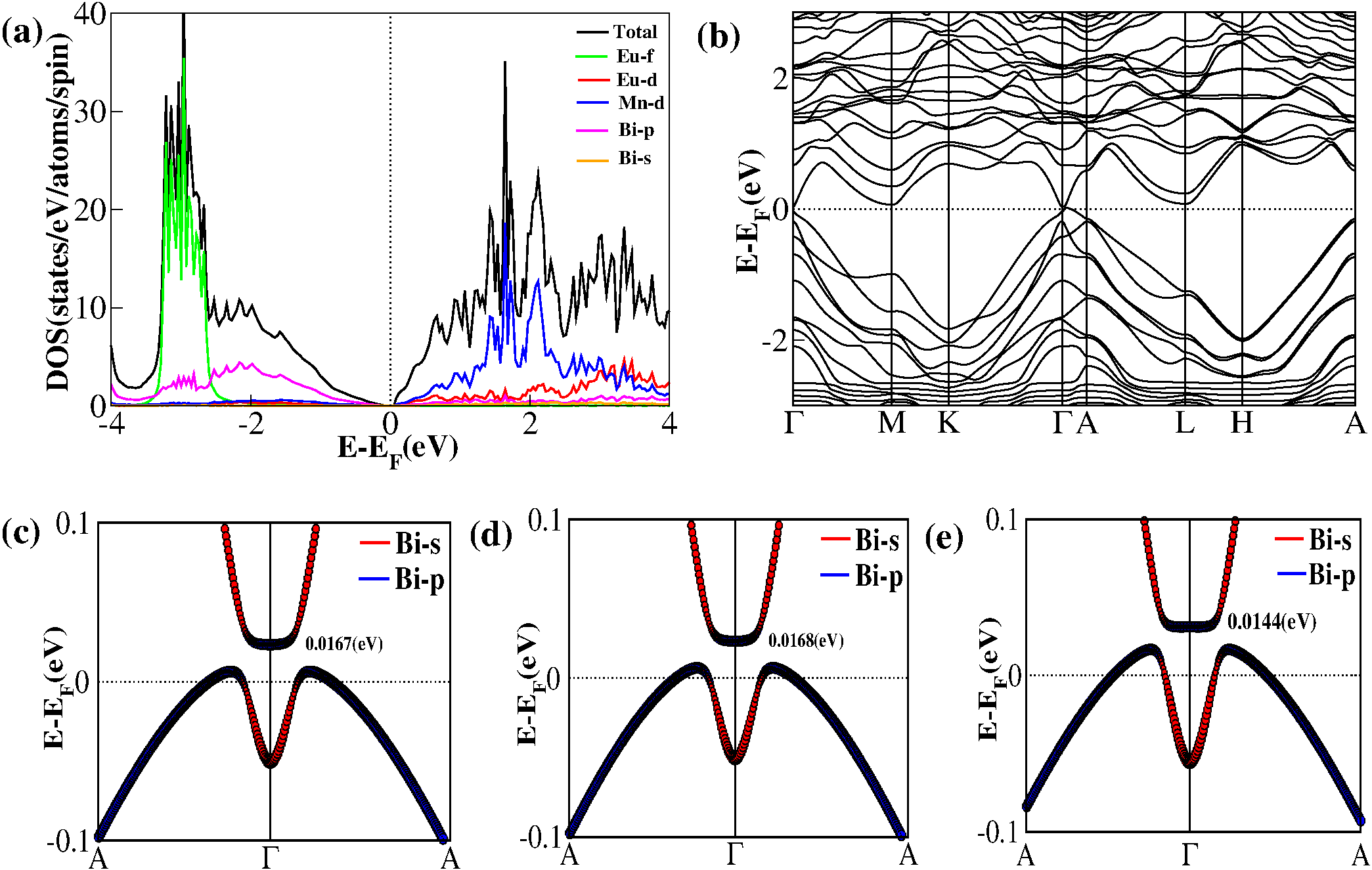}
	\end{center}
	\caption { \noindent (a) DOS and (b) band structure of EuMn$_2$Bi$_2$ calculated within GGA+U+SO approximation for C0-AFM3 magnetic configuration. Band structure along A$-$$\Gamma$$-$A direction with magnetic moments of Eu and Mn along the (c) [100] direction (C0-AFM2) (d) [110] direction (C0-AFM3) and (e) [111] direction (C0-AFM4) showing band inversion. }
	\label{Fig6}
	
\end{figure}
\begin{figure}
	\begin{center}
		\includegraphics[width=8.5cm]{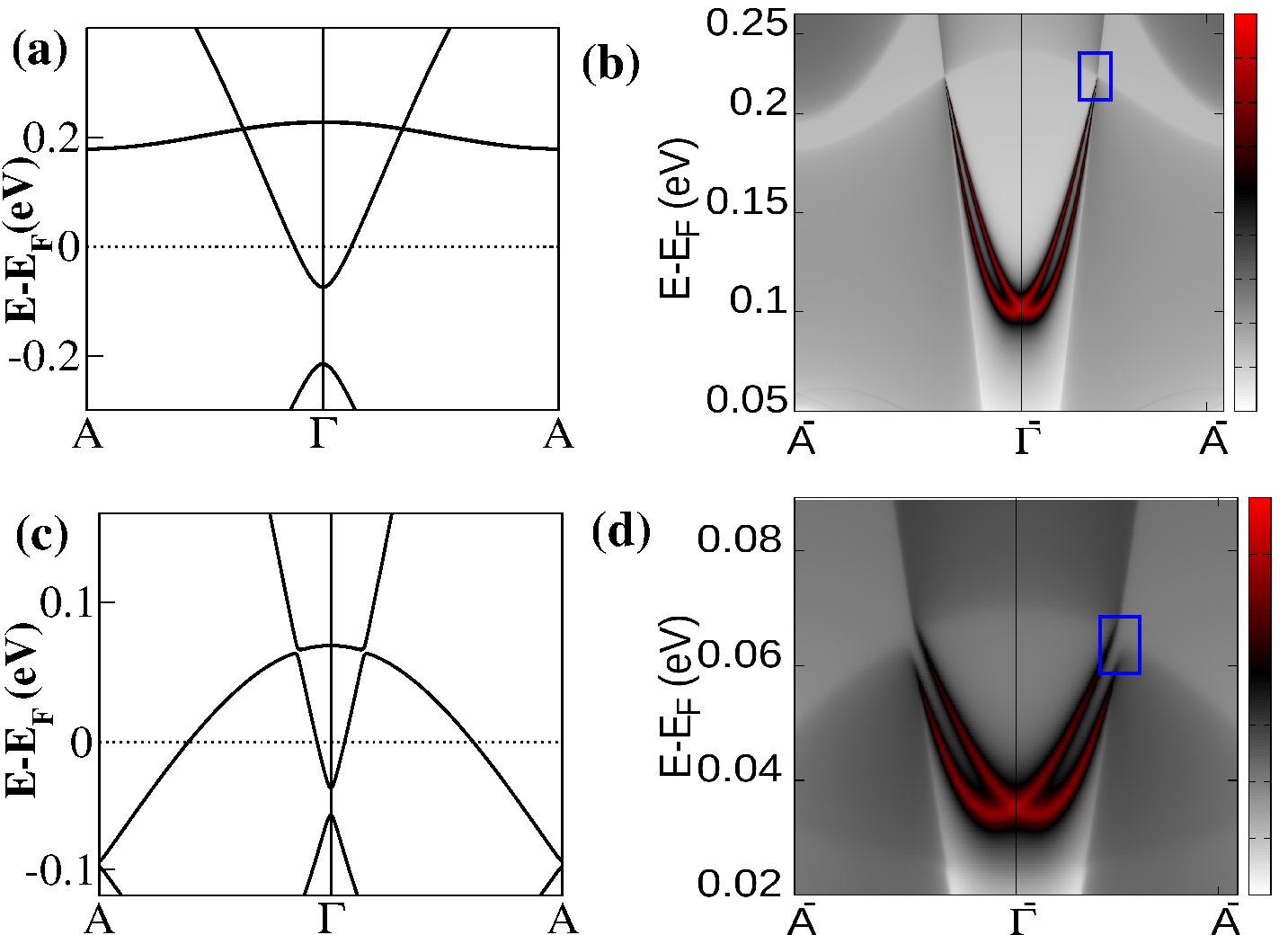}
	\end{center}
	\caption { \noindent  GGA+U+SO band structure of EuMn$_2$Bi$_2$ for  (a) G0-AFM1 and (c) G0-AFM2 magnetic configurations. (b) and (d) are the plots illustrating the spectral function \(A(\bar{\mathbf{k}},E)\) within a slab geometry, with \(\bar{\mathbf{k}}\) representing momentum within the surface Brillouin zone. Distinct red bands emerge in between the bulk continuum in these plots, unveiling surface state spectrum (b) the drumhead-like surface states in red near the $\bar{\Gamma}$ point connecting DPs on (100) surface (d) Topological insulator surface state spectrum on (100) surface. The colorbar scale is in arbitrary units.}
	\label{Fig7}
\end{figure}

{\it Electronic structure:}
We have computed the partial density of states (PDOS) and band structure of EuMn$_2$As$_2$ for the ground state magnetic configuration as discussed above within GGA+U+SO approximation with U$_{eff}$ values of 5 eV and 7 eV applied to Mn d and Eu f states respectively. The PDOS and band structure are presented in  Fig~\ref{Fig4}(a) and Fig~\ref{Fig4}(b) respectively. We observe an indirect band gap of 0.67 eV confirming the insulating ground state of this system observed experimentally.

\subsection{EuMn$_2$Bi$_2$}
{\it Dynamical Stability:}
Next, we delved into exploring the magnetic ground state and electronic structure of a new compound EuMn$_2$Bi$_2$. We started with the experimental structural data of EuMn$_2$As$_2$ and performed a full structural relaxation after substituting all As with Bi. We carried out phonon spectrum calculations to assess the dynamical stability of the structure. Remarkably, all phonon frequencies were found to be positive throughout the entire Brillouin zone, as illustrated in Fig~\ref{Fig5} where we presented the phonon band structure and density of states (DOS) in Fig~\ref{Fig5} (a) and (b) respectively, affirming the dynamical stability of the EuMn$_2$Bi$_2$ compound. 

{\it Magnetic ground state:}
Similar to EuMn$_2$As$_2$, we performed calculations to determine the magnetic ground state of the Mn sublattice in EuMn$_2$Bi$_2$ by retaining the 4f electrons of Eu in the core. Within GGA+U, the C0-AFM is found to be the ground state for range of U$_{eff}$ values (i.e. 0 eV, 3 eV, 5 eV, 7 eV and 11 eV) with the energy difference between the C0-AFM and G0-AFM being 12.76 meV, 9.3 meV, 6.46 meV, 4.5 meV, and 2.3 meV, respectively. These differences are notably smaller compared to the same in EuMn$_2$As$_2$. After incorporating spin-orbit coupling in the calculations within GGA+U+SO approximation, the magnetic moments of the Mn atoms are found to align between the crystallographic a and b axes [110] for the C0-AFM configuration, establishing it as the ground state exactly same as EuMn$_2$As$_2$. Lastly, upon considering the Eu 4f electrons as valence, Eu moments are found to order in A-AFM with Mn moments in C0-AFM. Our GGA+U+SO calculations performed for same 32 different configurations as shown in Table ~\ref{Tab4} (Appendix-A)  further reveals that in the ground state Eu/Mn moments point along [110] direction. We also see that when both Eu and Mn moments point along [100] direction, the state is energetically very close to the ground state. 

{\it Electronic structure:}
The DOS  and band structure for the magnetic ground state of EuMn$_2$Bi$_2$, calculated using the GGA+U+SO approximation are presented in Fig~\ref{Fig6}(a) and Fig~\ref{Fig6}(b) respectively. Comparing with the corresponding DOS of EuMn$_2$As$_2$ (Fig~\ref{Fig4}(a)) calculated using the same U values for Eu and Mn, one observes that the band-gap has substantially reduced in case of EuMn$_2$Bi$_2$. The conduction and valence band almost touch each other at $\Gamma$ point (see Fig~\ref{Fig6}(b)). Looking close into this region around $\Gamma$ point we observe that there exists a tiny band gap ($\sim$16 meV) with band inversion between Bi s and Bi p bands (see Fig~\ref{Fig6}(d)). However, we didn't observe any surface states. Therefore, we believe EuMn$_2$Bi$_2$ in it's magnetic ground state will be a normal insulator with a tiny band gap. 
\begin{figure}
	\begin{center}
		\includegraphics[width=8.5cm]{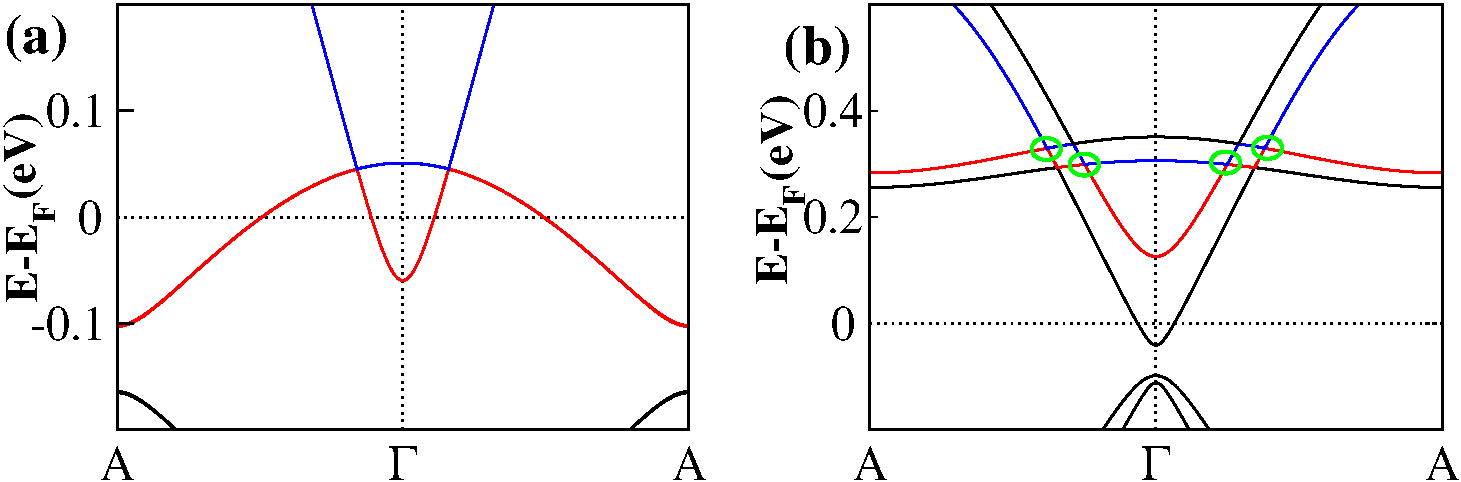}
		\includegraphics[width=8.5cm]{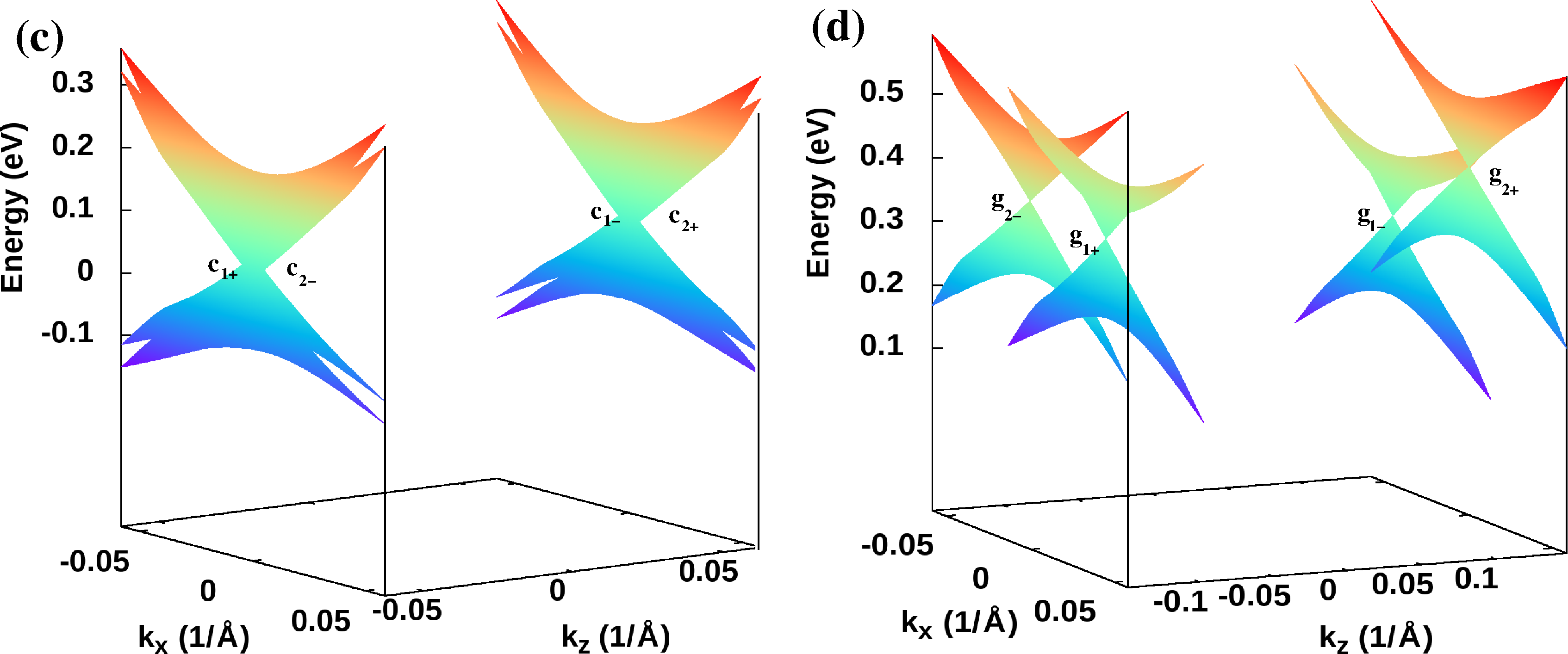} 
	\end{center}
	\caption { \noindent Band structure of EuMn$_2$Bi$_2$ calculated within GGA+U+SO for (a) C0-AFM1 and (b) G0-AFM5 magnetic configurations. The 3D band plots showing the WPs in (c) C0-AFM1 and (d) G0-AFM5 phases.}
	\label{Fig8}
\end{figure}
\begin{figure}
	\begin{center}
		\includegraphics[width=8.5cm]{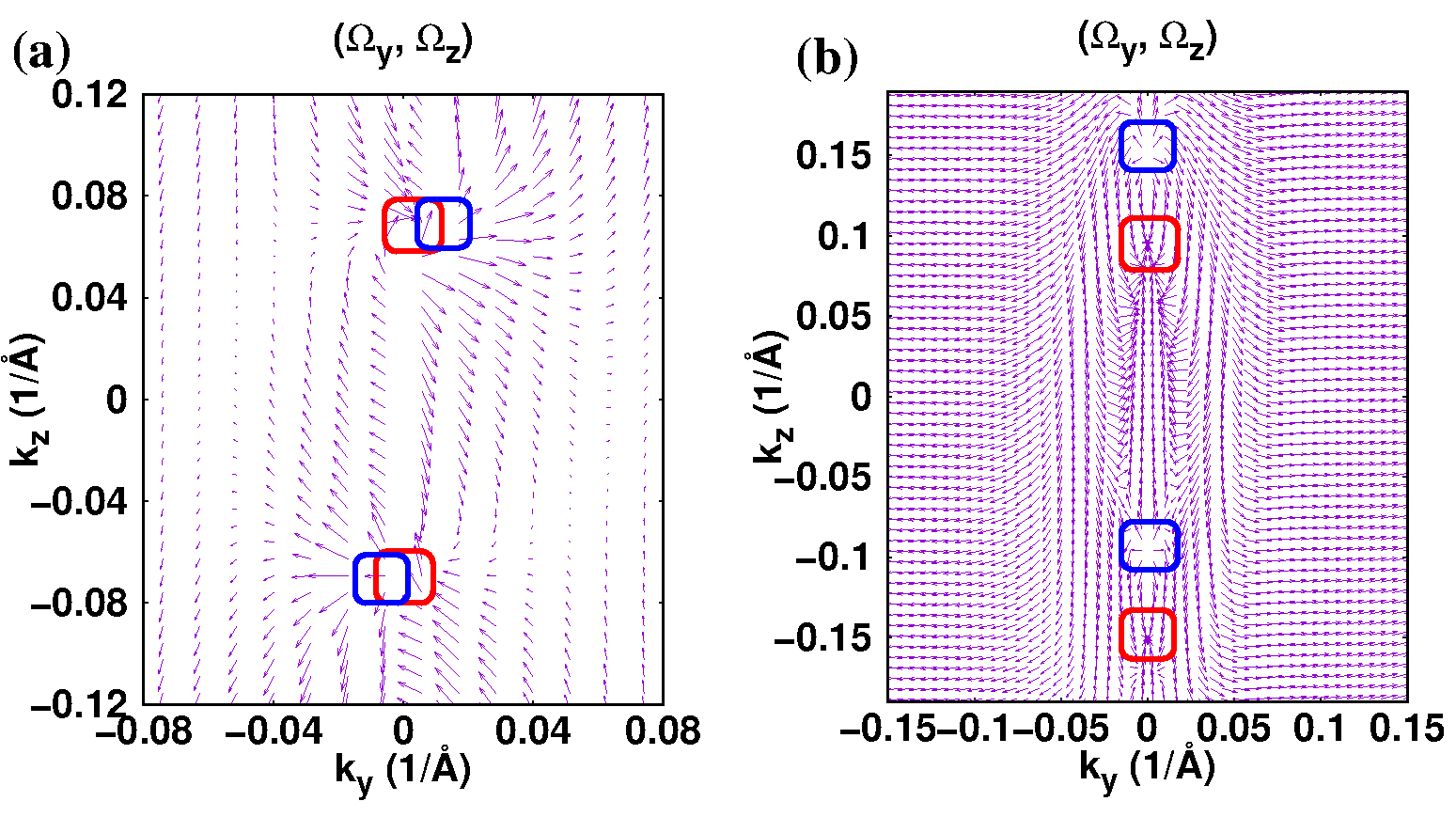}
	\end{center}
	\caption { \noindent The Berry curvature is depicted on the k$_y$-k$_z$ plane for the two sets of WPs , with the source (blue box) and sink (red box) identified for WPs in (a) C0-AFM1 and (b) G0-AFM5.}
	\label{Fig9}
\end{figure}

\begin{figure}
	\begin{center}
		\includegraphics[width=9.0cm]{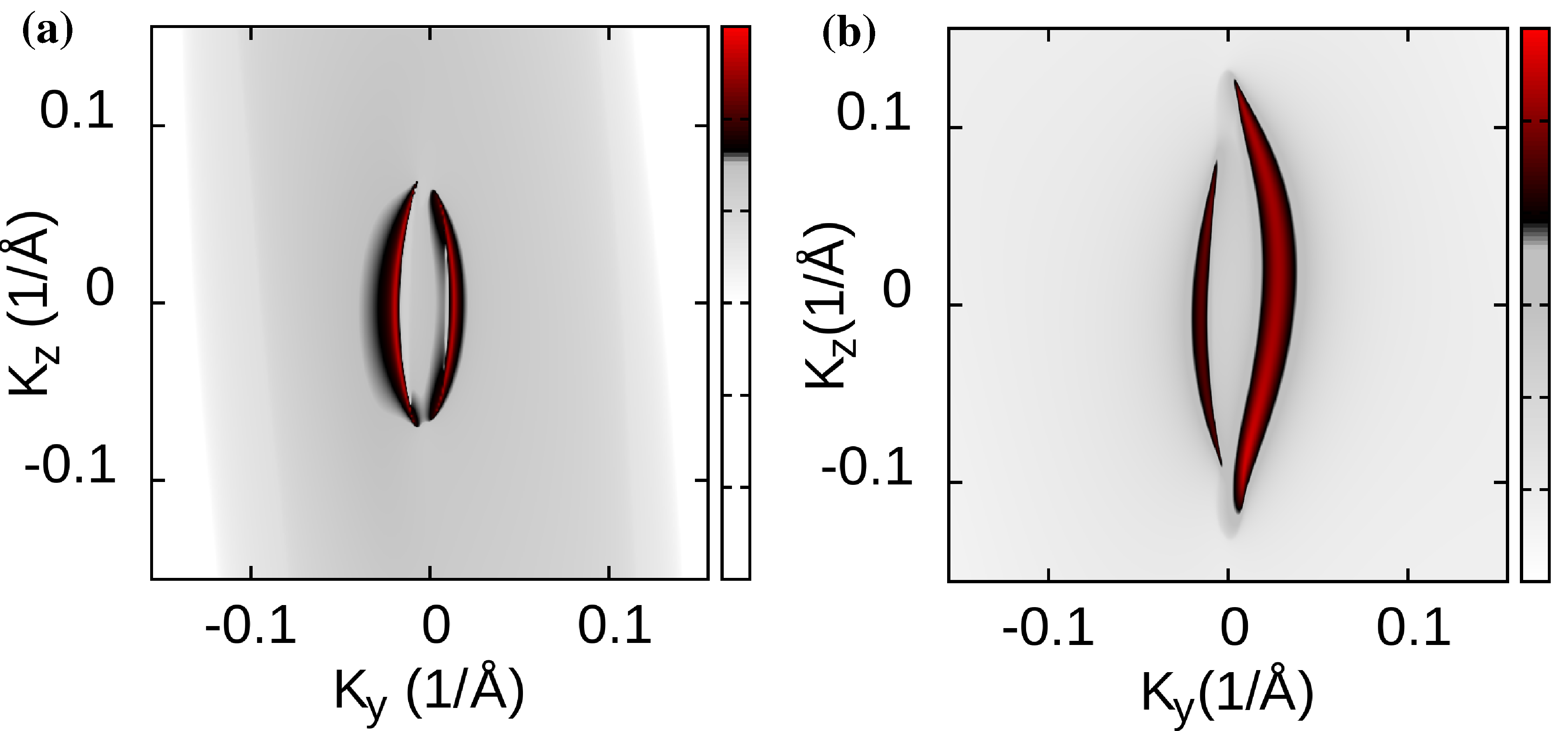}
		\caption{Displaying the two Fermi arcs on the \(k_y\)-\(k_z\) plane for the two sets of WPs , (a) C0-AFM1 and (b) G0-AFM5.The colorbar scale is in arbitrary units.	}
		\label{Fig10}
	\end{center}
\end{figure} 
{\it Topological properties:}
While the C0-AFM3 magnetic order of EuMn$_2$Bi$_2$  exhibits greater stability compared to 32 other magnetic configurations, the energy differences are minimal, typically around 10 meV or even less in some cases. Recent investigations have shown that in the case of hexagonal Eu compounds like EuCd$_2$As$_2$, which contains only one magnetic element (Eu), the manipulation of magnetic order is possible by adjusting growth conditions to produce either AFM or FM \cite{soh,ECAmani1,ECAmani2,ECAmani3}. Some theoretical reports also explain that this change in the magnetic order can be achieved with doping or external pressure\cite{AA,preECA,dopECA,tm}. Consequently, our analysis focuses on the examination of the topological properties associated with various magnetic configurations which are energetically close to each other as well as to the ground state as discussed below.

{\it Dirac semimetal and topological insulator:}
First we consider the G0-AFM1 and G0-AFM2 configurations as shown in Fig~\ref{Fig3}(f), Fig~\ref{Fig3}(g) respectively. Looking closely along $\Gamma$-A direction reveals very interesting topological properties.  When the Eu/Mn moments point along $z$-direction we see band crossings slightly away from $\Gamma$ point (see Fig~\ref{Fig7}(a)). We identify these crossings as Dirac points (DPs) as we describe below. The inversion symmetry (P) is present in this system. In the G0-AFM1 configuration, the intrinsic magnetism destroys the Time-Reversal Symmetry (TRS). However, a non-symmorphic time-reversal symmetry denoted as ${\cal S}\!=\!{\cal T} \oplus c$ persists. This symmetry connects the two layers with opposite spins at $z\!=\!0$ and $z\!=\!c$, ensuring the anti-unitarity of the combined operation ${\cal SP}$, despite the explicit breaking of TRS. In addition, G0-AFM1 magnetic state also possesses $C_{3z}$ rotational symmetry along with ${\cal SP}$ due to which we observe the doubly degenerate bands crossing at DPs along $\Gamma$-A direction\cite{nagaosa}. These DPs are connected by two drumhead-like Fermi arcs which we observe through the surface states on the (100) surface, as shown in Fig~\ref{Fig7}(b). Interestingly, when the Eu/Mn moments point away from z-direction, the $C_{3z}$ symmetry is broken. Consequently, a band gap emerges, as one can see in the Fig~\ref{Fig7}(c). The corresponding surface state spectrum on the (100) surface (presented in Fig~\ref{Fig7}(d)) clearly shows the crossing between the surface states while the bulk bands are gapped. This strongly indicates that the system becomes a topological insulator from Dirac semimetal when the moments deviate from $z$ direction in G0-AFM phase. Similarly, G0-AFM3, G0-AFM4 will also be topological insulators. 

{\it Weyl semimetal and normal insulator :}
In our subsequent analysis, we explore the C0-AFM1 and G0-AFM5 configurations, pictorially represented in Fig~\ref{Fig3}(a) and Fig~\ref{Fig3}(j), respectively. Looking closely at the band dispersion along $\Gamma$-A direction (see Fig~\ref{Fig8}(a) and Fig~\ref{Fig8}(b)) we find very interesting band toplogy. 
As we move from G0-AFM1 to C0-AFM1 or G0-AFM5, the non-symmorphic time-reversal symmetry (${\cal S}$) is broken, leading to the lifting of band degeneracy in the $\Gamma$-A direction. But due to the preservation of $C_{3z}$ symmetry (as Eu/Mn both moments are pointed along $z\!$ axis) and inversion symmetry (P), the non-degenerate bands cross each other to form Weyl points (WPs). Consequently, each pair of DPs in the $\Gamma$-A direction (present in G0-AFM1) undergoes splitting, resulting in four pairs of WPs in C0-AFM1 and G0-AFM5 configuration. The specific coordinates of the four WPs and their chirality are summarized in Table ~\ref{Tab2}. In case of C0-AFM1 we observe two WPs on either side of $\Gamma$ point very close to each other (see Fig~\ref{Fig8}(c)). They can be clearly seen in the Berry curvature plot presented in Fig~\ref{Fig9}(a). Whereas, in case of G0-AFM5 the WPs are clearly separated as we can see from the band structure presented in  Fig~\ref{Fig8}(b) and Fig~\ref{Fig8}(d). The Berry curvature plot presented in Fig~\ref{Fig9}(b) also shows their locations. Another prominent characteristic of Weyl semimetal is the presence of Fermi arcs that connect WPs with opposite chirality on the surface. The chirality of the WPs is distinctly evident through the analysis of Berry curvature for both the cases presented in Fig~\ref{Fig9}(a) and (b). The presence of two distinct Fermi arcs connecting each pair of WPs (see Fig~\ref{Fig10}(a) and (b)) confirms that EuMn$_2$Bi$_2$ becomes a Weyl semimetal in its C0-AFM1 and G0-AFM5 phases. Same will be the case of C0-AFM5. However, when Eu/Mn moments deviate from the $z\!$ direction (e.g. C0-AFM2, C0-AFM3 and C0-AFM4), a band gap emerges due to the breaking of the ($C_{3z}$) symmetry, rendering it a normal insulator (Fig~\ref{Fig6}(c)-(e)). Inspite of the presence of the band inversion, the absence of topological surface states classifies them as normal insulators.
\section{Conclusions}
Using first principles DFT calculations we have performed a thorough study of the nature of two magnetic transitions and the magnetic ground state of EuMn$_2$As$_2$ which has been previously studied experimentally\cite{anand, dahal}. Our calculations reveal that in the absence of Eu moments ordering, Mn sublattice prefers to order in C0-AFM state with Mn moments residing in the $ab$-plane. Corresponding Mn-Mn exchange interaction values explain the high temperature magnetic transition seen in experiments due to Mn moments' ordering. Eu moments, on the other hand, are observed to order in A-AFM again preferring to lie in $ab$-plane. However, a low lying excited state with both Eu/Mn moments oriented out of $ab$-plane along [111] direction is also observed indicating possibility of spin-reorientation with slight perturbation. Eu-Eu exchange interaction values are observed to be an order of magnitude less than the corresponding Mn-Mn exchange values indicating Eu moments order at much lower temperatures as observed in experiments. Our electronic structure calculations of the magnetic ground state within GGA+U+SO approximations establishes the insulating nature of the compound. Interestingly, when we replaced As by Bi in EuMn$_2$As$_2$ to create a new compound EuMn$_2$Bi$_2$, we observed that it is dynamically stable with no soft phonon modes indicating the feasibility of it's experimental preparation in the laboratory. Similar to EuMn$_2$As$_2$, we found that EuMn$_2$Bi$_2$ also should have two magnetic transitions with Mn moments ordering in C0-AFM at higher temperature and Eu moments ordering in A-AFM at lower temperature. Further, we observe that there are several magnetically ordered state with C0-AFM and G0-AFM configurations having competing energies. These magnetic configurations show remarkable band topology. From our detailed analysis of band structure, Berry curvature, surface states etc. we conclude that EuMn$_2$Bi$_2$ in it's C0-AFM1 and G0-AFM5 phase would be a Weyl semimetal while in G0-AFM1 phase it is a Dirac semimetal and finally the G0-AFM2 phase is a topological insulator and so on. Within G0-AFM phase one can go from Dirac semimetal to topological insulator state by moving the Eu/Mn spin orientations away from $z$ direction. Similarly, within C0-AFM phase one can go from Weyl semimetal to normal insulator state by moving the Eu/Mn spin orientations away from $z$ direction. Therefore, EuMn$_2$Bi$_2$ will be a very verstile candidate material with a range of magnetic order driven topological states which can be tuned from one to the other by external handle. Future experiments will shed more light on this new compound.

\section{Acknowledgements} 

TM acknowledges Science \& Engineering Research Board (SERB), India for funding through SERB-POWER research grant (no. SPG/2021/000443). AC and TM acknowledge the National Supercomputing Mission (NSM) for providing computing resources of ‘PARAM Ganga’ at the Indian Institute of Technology Roorkee, which is implemented by C-DAC and supported by the Ministry of Electronics and Information Technology (MeitY) and Department of Science and Technology (DST), Government of India. AC acknowledges Ministry of Education, India for research fellowship.

\section{Appendix A}
\begin{table*}
	\begin{center}
		\caption{The total energy per formula unit is calculated in meV, with respect to the ground state, considering various orientations of Eu and Mn moments within the GGA+U+SO approximations for EuMn$_2$As$_2$. Here Eu/Mn moment direction is denoted by the $\mu$.}
		
		\label{Tab3}
		\begin{tabular}{m{2.0cm}  m{2.0cm} m{1.5cm} m{1.5cm}  m{1.5cm} m{1.5cm} || m{1.5cm} m{1.5cm}  m{1.5cm} m{1.5cm}}
			\hline\hline	\\
			{\bf GGA+U+SO}& & & {\bf C0-AFM} & & & & {\bf G0-AFM}  \\\\
			& & &  {\bf Mn ($\mu$)}& & & & & {\bf Mn($\mu$)} & \\
			& & {\bf[100]} & {\bf [001]} & {\bf[110]} & {\bf[111]} &{\bf [100]} & {\bf[001]} & {\bf[110] }& {\bf[111]}\\\\
			\hline \\
			& {\bf[100] }					& 0.13 & 7.33 & 7.39 & 5.15 & 16.81 & 20.05 & 18.91 & 19.52   \\\\
			
			{\bf Eu ($\mu$)} & {\bf0.0.1 }   & 7.23 & 0.51 & 4.08 & 5.21 & 19.82 & 17.14 & 20.04 & 19.56\\\\

			&{\bf[110]}                    & 4.16 & 7.32 & 0.00 & 3.01 & 18.84 & 20.04 & 16.75 & 18.31	\\\
			
			&{\bf[111] }                   & 5.10 & 5.27 & 2.99 & 0.04 & 19.34 & 19.53 & 16.79 & 18.24 \\\\
			
			\hline
		\end{tabular}
		
	\end{center}
\end{table*}
\begin{table*}
	\begin{center}
		\caption{The total energy per formula unit is calculated in meV, with respect to the ground state, considering various orientations of Eu and Mn moments within the GGA+U+SO approximations for EuMn$_2$Bi$_2$. Here Eu/Mn moment direction is denoted by the $\mu$.}
		\label{Tab4}
		\begin{tabular}{m{2.0cm}  m{2.0cm} m{1.5cm} m{1.5cm}  m{1.5cm} m{1.5cm} || m{1.5cm} m{1.5cm}  m{1.5cm} m{1.5cm}}
			\hline\hline	\\
			{\bf GGA+U+SO}& & & {\bf C0-AFM} & & & & {\bf G0-AFM}  \\\\
			& & &  {\bf Mn ($\mu$)}& & & & & {\bf Mn($\mu$)} & \\
			& & {\bf[100]} & {\bf[001]} & {\bf[110]} & {\bf[111]} &{\bf [100]} & {\bf[001]} & {\bf[110] }& {\bf[111]}\\\\
			\hline \\
			& {\bf[100] }					& 0.01 & 7.16 & 2.81 & 4.25 & 12.42 & 11.49 & 13.4 & 13.58   \\\\
			
			{\bf Eu ($\mu$)} & {\bf[001] }   & 5.12 & 2.39 & 5.09 & 4.25 & 11.17 & 14.17 & 11.16 & 13.89 \\\\
			
			&{\bf[110] }                   & 2.79 & 2.79 & 0.00 & 2.58 & 13.46 & 11.58 & 12.45 & 13.45 \\\\
			
			&{\bf[111]}                    & 3.55 & 5.61 & 1.94 & 0.69 & 13.44 & 14.45 & 13.27 & 12.93	\\\\
			\hline
		\end{tabular}
		
	\end{center}
\end{table*}

\end{document}